\documentclass[journal]{IEEEtran}

\usepackage{iftex}
\ifXeTeX % Times-compatible font for XeTeX/Tectonic builds; pdflatex (arXiv) uses IEEEtran's default Times
  \usepackage{fontspec}
\fi
\usepackage{cite}
\usepackage{amsmath,amssymb}
\usepackage{graphicx}
\usepackage{booktabs}
\usepackage{array}
\usepackage[hidelinks]{hyperref}

\newcommand{\figw}{0.8\columnwidth}

\begin{document}

\title{A Secure dToF LiDAR SoC with Dual-Domain Fingerprinting and Event-Driven AFE Circuit Achieving Sensor-Level Attack Resilience}

\author{Risa~Nonaka$^{*}$,
        Ryoya~Matsuno$^{*}$,
        Shota~Nagai$^{*}$,
        Satomi~Miyagi$^{*}$,
        Yuki~Hayakawa,
        Ryo~Suzuki,
        Kazuma~Ikeda,
        Ozora~Sako,
        Rokuto~Nagata,
        Ryo~Yoshida,
        Shimpei~Ando,
        Wenlun~Zhang,
        and~Kentaro~Yoshioka
\thanks{All authors are with Keio University, Japan.
Corresponding author: Kentaro Yoshioka (e-mail: kyoshioka47@keio.jp).}%
\thanks{$^{*}$R. Nonaka, R. Matsuno, S. Nagai, and S. Miyagi contributed equally to this work.}}

\maketitle

\begin{abstract}
Recent studies have shown that most commercial direct time-of-flight (dToF) LiDARs can be spoofed by injecting high-frequency laser pulses into the receiver, which can erase pedestrians from the point cloud. This paper presents the first dToF LiDAR system-on-chip (SoC) with integrated sensor-level hardware security against spoofing attacks. We propose Dual-Domain Fingerprinting (DDF), which emits laser pulse pairs whose time interval and amplitude ratio are both randomized and authenticates received echoes in this two-dimensional space, so that spoofed signals are rejected before they corrupt the ranging result. An Event-Driven AFE (ED-AFE) activates the ADC only around pulse peaks: it digitizes three samples around each peak with a triggered ADC at 1-GHz sampling and applies parabolic interpolation, achieving 1-cm distance resolution with a 99\% reduction in ADC power. A time-modulated laser driver controls the laser amplitude from 10\% to 100\% by modulating the charge time of a laser capacitor, providing the microsecond-order amplitude modulation required by DDF. A LiDAR system with a 16-channel 65-nm CMOS prototype SoC demonstrates up to 120-m ranging and an AFE power of 3.1~mW per channel, 60\% lower than prior art. In a proof-of-concept experiment in which the dual-domain authentication is applied to measured sensor data, 73\% of the point cloud is protected under spoofing attack, compared with 0\% without DDF.
\end{abstract}


% ===========================================================================
\section{Introduction}
\label{sec:intro}

\begin{figure}[!t]
\centering
\includegraphics[width=\figw]{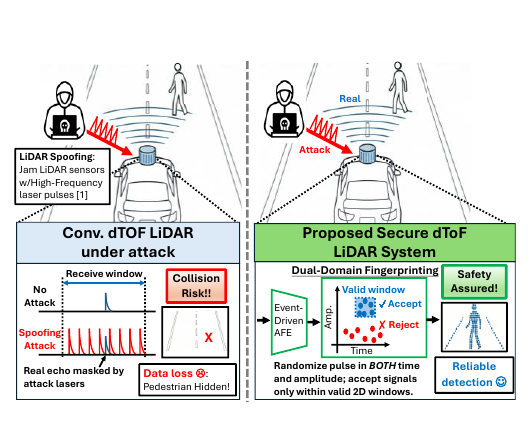}
\caption{Comparison between a conventional dToF LiDAR and the proposed secure dToF LiDAR system under spoofing attacks.}
\label{fig:concept}
\end{figure}

\IEEEPARstart{L}{iDAR} sensors are indispensable for Level-4+ autonomous driving, providing real-time 3-D perception for safe navigation. However, recent studies \cite{Sato_NDSS2024, Sato_NDSS2025} reveal that most commercially available LiDAR sensors are fundamentally vulnerable to spoofing attacks, in which injected high-frequency malicious laser pulses corrupt dToF measurements. Such an attack can erase pedestrians from the point cloud, leading to catastrophic safety failures in autonomous driving (Fig.~\ref{fig:concept}).

Despite this critical threat, most prior LiDAR research has focused on improving range, resolution, and scalability \cite{Yoshioka_ISSCC2018, Kondo_ISSCC2020, Zou_ISSCC2024}, leaving hardware-level LiDAR security largely unexplored. Software-based attack detection \cite{Suzuki_Sensors2025} cannot prevent corrupted measurements from being generated at the sensor output. Timing-only pulse fingerprinting, adopted in commercial products \cite{Hesai_Pandar128}, has been shown to be defeated by adaptive attacks \cite{Sato_NDSS2025}. Other LiDAR AFEs, including pulse-coded \cite{Kashmiri_ISSCC2020} and multiplexed ADC/TDC \cite{Liu_TCAS2020} designs, likewise do not address attack resilience. While FMCW LiDAR offers some inherent robustness, its high system cost limits adoption in cost-sensitive automotive markets. These limitations highlight the need for a fundamentally new, sensor-level hardware approach to secure dToF LiDAR operation.

\begin{figure}[!t]
\centering
\includegraphics[width=\figw]{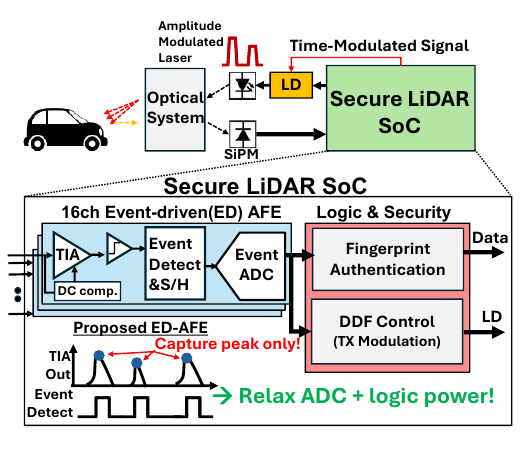}
\caption{Overall architecture of the proposed secure dToF LiDAR SoC.}
\label{fig:arch}
\end{figure}

In this work, we introduce a secure LiDAR system featuring Dual-Domain Fingerprinting (DDF) (Fig.~\ref{fig:arch}). By embedding authentication in both the laser amplitude and timing domains, the proposed approach defends against LiDAR spoofing attacks. The system is built on three key circuit techniques: (1) a DDF-based authentication system, (2) an Event-Driven AFE (ED-AFE), and (3) a time-modulated laser driver. The ED-AFE significantly reduces ADC power consumption by activating the ADC only around the laser peaks, achieving a 99\% reduction in ADC power. The proof-of-concept LiDAR system in 65-nm CMOS maintains a ranging distance of up to 120~m and, with DDF applied to measured sensor data, protects 73\% of the point cloud under spoofing attack. In addition, the proposed ED-AFE reduces the per-channel AFE power consumption by 60\% compared with prior art.

% ===========================================================================
\section{Related Work}
\label{sec:related}

\textit{LiDAR receiver circuits:} For scanning LiDARs with off-chip photodetectors, ADC-based, hybrid TDC/ADC, dual-data-converter, and multiplexed ADC/TDC AFEs have been reported \cite{Yoshioka_JSSC2018, Yoshioka_ISSCC2018, Kondo_ISSCC2020, Liu_TCAS2020}; current-integrating TIAs avoid a high-speed ADC at the expense of a 10-GHz clock \cite{Zou_ISSCC2024}; and pulse-coded front ends recover echoes by pulse compression \cite{Kashmiri_ISSCC2020}. SPAD-based dToF sensors integrate the detector with timing circuits \cite{Niclass_JSSC2014}, including 3-D-stacked imagers with in-pixel histogramming \cite{Hutchings_JSSC2019} and automotive stacked SPAD sensors \cite{Kumagai_ISSCC2021, Padmanabhan_ISSCC2021}, and some suppress interference from other light sources \cite{Ximenes_ISSCC2018, Seo_JSSC2021}. These designs target range, resolution, and channel count, and do not authenticate received pulses against a deliberate attacker.

\textit{Physical attacks on LiDAR:} Early remote attacks relayed or replayed laser pulses to create fake objects or blind the sensor \cite{Petit_BH2015, Shin_CHES2017}. Later work crafted spoofed points that deceive 3-D object detectors \cite{Cao_CCS2019, Sun_USENIX2020}, removed real objects from the perceived scene \cite{Hau_AutoSec2021, Cao_USENIX2023, Jin_SP2023}, and attacked multi-sensor fusion \cite{Cao_SP2021, Hallyburton_USENIX2022}. Sato \emph{et al.} showed that high-frequency spoofing removes objects from most tested new-generation LiDARs \cite{Sato_NDSS2024}; follow-up studies demonstrated attacks against moving vehicles \cite{Suzuki_VehicleSec2024} and an adaptive attack that defeats pulse-fingerprinted LiDARs at high speed and long distance \cite{Sato_NDSS2025}.

\textit{Defenses:} Perception-level defenses detect spoofed points after the point cloud is formed, using occlusion patterns \cite{Sun_USENIX2020}, shadow consistency \cite{Hau_ESORICS2021}, or temporal consistency \cite{Cho_BMVC2023}; a recent detector uses the frequency-domain signature of raw ToF data \cite{Suzuki_Sensors2025}. Such detectors can flag an attack but cannot restore a real echo that has been replaced by attack pulses. Pseudo-random pulse timing has been used to reject pulses from other LiDARs \cite{Kim_RS2022}, and commercial LiDARs randomize the pulse interval \cite{Hesai_Pandar128}; however, timing-only fingerprints can be defeated by a deliberate attacker \cite{Sato_NDSS2025}. Physical-layer authentication has been explored with an AES-based fingerprint superimposed on the emitted light \cite{Matsumura_CANDARW2018} and with Doppler-shift-based sender verification \cite{Hu_IoTJ2024}. Coherent FMCW LiDAR provides inherent selectivity against incoherent light \cite{Behroozpour_JSSC2017, Poulton_OL2017, Rogers_Nature2021}, but its complexity and cost remain a barrier for automotive applications. Finally, event-driven and level-crossing acquisition reduces conversion energy for sparse signals \cite{Tsividis_TCASII2010}, of which LiDAR echoes are an extreme case.

In contrast to the above, this work adds a second, randomized amplitude domain to the pulse fingerprint and is designed to perform the authentication in the receiver hardware, supported by an event-driven AFE that provides the required pulse amplitudes at low power.

% ===========================================================================
\section{Attack and Threat Model}
\label{sec:threat}

A dToF LiDAR emits a laser pulse and measures the round-trip time of flight $t_{\mathrm{ToF}}$ of the echo; the target distance is $d = c\,t_{\mathrm{ToF}}/2$, where $c$ is the speed of light. Because any light arriving in the receive window is indistinguishable from a legitimate echo, an attacker can corrupt this measurement. In the high-frequency spoofing attack of \cite{Sato_NDSS2024}, the attacker floods the receiver with high-frequency laser pulses; the real echo is masked by the attack pulses, and the corresponding point is lost (Fig.~\ref{fig:concept}, left). Adaptive variants remain effective against LiDARs that employ timing fingerprints \cite{Sato_NDSS2025}.

We consider an external attacker who injects laser pulse trains into the LiDAR receiver with arbitrary timing and intensity, and who may know the structure of the defense, but cannot predict the random parameters that the victim LiDAR selects for each shot. The goal of the attacker is to corrupt the ranging result, e.g., to erase a pedestrian from the point cloud; the goal of the defense is to use only echoes of the LiDAR's own laser for ranging.

% ===========================================================================
\section{Dual-Domain Fingerprinting}
\label{sec:ddf}

\subsection{Limitation of Time-Domain Fingerprinting}

\begin{figure}[!t]
\centering
\includegraphics[width=\figw]{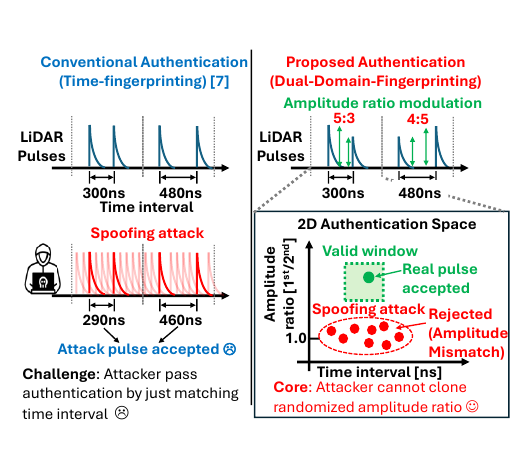}
\caption{Principle of the proposed Dual-Domain Fingerprint authentication.}
\label{fig:ddf}
\end{figure}

Conventional dToF LiDAR systems, which measure ToF using a single laser pulse, are inherently vulnerable to external interference and malicious spoofing. To mitigate this, time-interval-based fingerprinting has been proposed (Fig.~\ref{fig:ddf}, left) \cite{Hesai_Pandar128}: the LiDAR emits pulse pairs whose interval $\Delta t$ is randomized (e.g., 300~ns and 480~ns), and the receiver accepts only pulse pairs with the expected interval. However, this scheme remains susceptible to spoofing attacks. When an attacker injects high-frequency laser pulses, statistically matching fake pulse pairs can be generated (e.g., 290~ns and 460~ns in Fig.~\ref{fig:ddf}), allowing the authentication mechanism to be bypassed by simply matching the time interval.

\subsection{Principle of DDF}

To overcome this limitation, we introduce DDF authentication as our core technology for secure LiDAR sensing. As shown in Fig.~\ref{fig:ddf} (right), for each shot $k$ the system randomly generates a laser pulse pair characterized by both a time interval $\Delta t_k$ and an amplitude ratio $r_k = A_{1,k}/A_{2,k}$ (e.g., 5:3 and 4:5), where $A_{1,k}$ and $A_{2,k}$ are the amplitudes of the first and second pulses. This creates a two-dimensional fingerprint space, in which authentication is performed by validating the received echoes against the expected parameters. Formally, a pair of detected pulses with measured interval $\Delta\hat{t}$ and amplitude ratio $\hat{r}$ is accepted only if it falls inside a valid window around the expected fingerprint, which can be written as
\begin{equation}
\left|\Delta\hat{t} - \Delta t_k\right| \leq \tau
\quad\text{and}\quad
\left|\hat{r} - r_k\right| \leq \epsilon,
\label{eq:accept}
\end{equation}
where $\tau$ and $\epsilon$ are the time and amplitude-ratio tolerances. Attack pulses that match only the time interval are rejected by the amplitude mismatch.

Because both pulses of a pair travel the same optical path and are reflected by the same target within hundreds of nanoseconds, a common attenuation factor cancels in the ratio, so that, as long as the receiver responds linearly and is not saturated, the amplitude ratio of the legitimate echo can be compared with the transmitted ratio. For an attacker, simultaneously matching both the time interval and the amplitude ratio becomes very difficult within a practical attack time window, which prevents the authentication bypass that is possible with timing-only fingerprinting.

% ===========================================================================
\section{Event-Driven AFE}
\label{sec:edafe}

\subsection{Motivation}

\begin{figure}[!t]
\centering
\includegraphics[width=\figw]{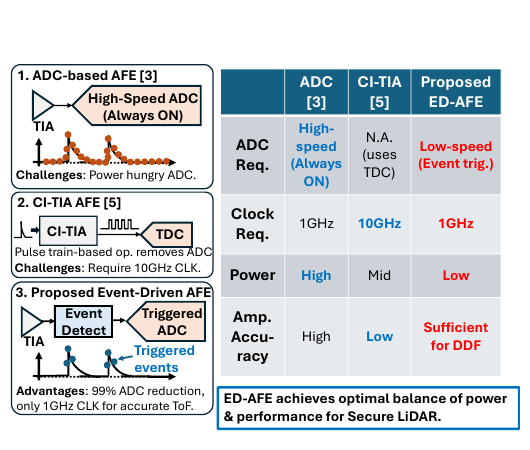}
\caption{Conceptual comparison between conventional LiDAR AFEs and the proposed event-driven AFE (ED-AFE).}
\label{fig:afecomp}
\end{figure}

DDF requires the AFE to measure the amplitude as well as the arrival time of the received pulses. Conventional LiDAR systems suffer from power inefficiency due to high-speed continuous ADCs \cite{Yoshioka_ISSCC2018} and CI-TIAs operating with 10-GHz clocks \cite{Zou_ISSCC2024}, the latter also offering limited amplitude accuracy (Fig.~\ref{fig:afecomp}). However, laser pulses meaningful for ranging are received for less than 1\% of the total time, making continuous monitoring highly power inefficient. Our proposed ED-AFE addresses this by operating intermittently, activating the ADC only when a pulse event is detected. The ED-AFE detects and digitizes only three voltage samples around the pulse peak sampled at 1-ns intervals, and converts them with a triggered ADC, which provides sufficient amplitude accuracy for DDF with low power.

\subsection{Architecture and Operation}

\begin{figure}[!t]
\centering
\includegraphics[width=\figw]{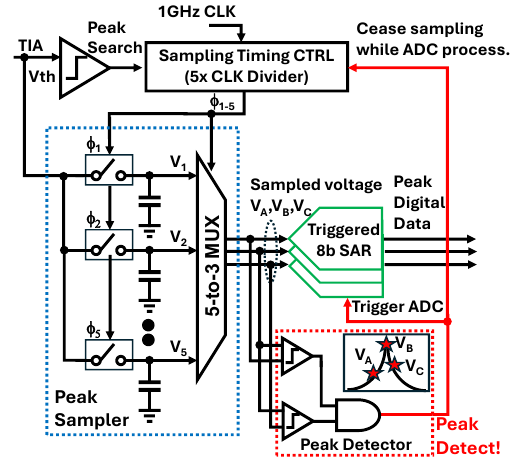}
\caption{Block diagram of the proposed event-driven LiDAR AFE.}
\label{fig:edafe_block}
\end{figure}

Fig.~\ref{fig:edafe_block} shows the block diagram of the ED-AFE. The TIA output is compared with a threshold $V_{\mathrm{th}}$ (peak search). A sampling timing controller ($5\times$ clock divider of the 1-GHz clock) generates phases $\phi_1$--$\phi_5$ for five samplers, a 5-to-3 MUX selects three samples ($V_A$, $V_B$, $V_C$), and a triggered 8-bit SAR ADC converts them into the peak digital data. Sampling is ceased while the ADC is processing.

\begin{figure}[!t]
\centering
\includegraphics[width=\figw]{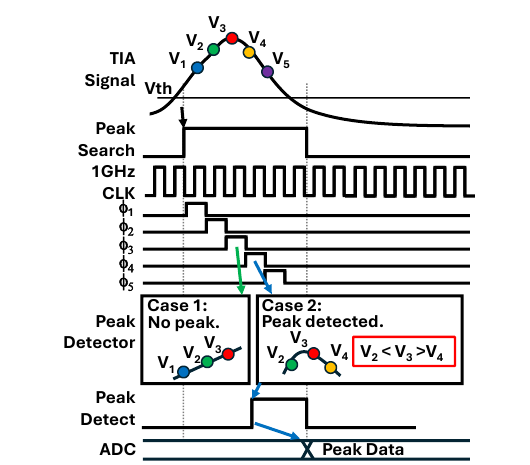}
\caption{Timing diagram of the proposed event-driven LiDAR AFE.}
\label{fig:edafe_timing}
\end{figure}

The timing diagram is shown in Fig.~\ref{fig:edafe_timing}. When the input voltage exceeds the predefined threshold, the TIA output is sequentially sampled and held by the five samplers at 1-GHz intervals. A peak detector then examines three consecutive voltage samples to determine the local signal trend. If all three samples exhibit a monotonic increase (Case~1), the signal is judged to be in the rising portion of a pulse, and sampling is continued. In contrast, if the three consecutive samples exhibit a local maximum ($V_2 < V_3 > V_4$, Case~2), the signal is identified as a pulse peak, at which point sampling is frozen and the ADC is triggered. Due to discrete-time sampling, the true peak may be missed; the ED-AFE therefore also triggers a peak event when three consecutive samples show a monotonic decrease, indicating a passed peak. By operating in this intermittent, event-driven manner, the ED-AFE significantly reduces the number of ADC activations, achieving a 99\% reduction in ADC power.

\subsection{Sub-Sample Peak Estimation}

With 1-GHz sampling alone, the timing resolution would be limited to 1~ns, i.e., 15~cm in distance. The ED-AFE applies parabolic interpolation to the three digitized samples. With $V_A$, $V_B$, and $V_C$ the three samples bracketing the peak, taken at $t_B - T_s$, $t_B$, and $t_B + T_s$ with $T_s$ the sampling period, the estimated peak time is
\begin{equation}
\hat{t} = t_B + \frac{V_A - V_C}{2\left(V_A - 2V_B + V_C\right)}\,T_s .
\label{eq:interp}
\end{equation}
This allows the ED-AFE to achieve 1-cm distance resolution while avoiding continuous high-speed conversion.

% ===========================================================================
\section{Time-Modulated Laser Driver}
\label{sec:laser}

\begin{figure}[!t]
\centering
\includegraphics[width=\figw]{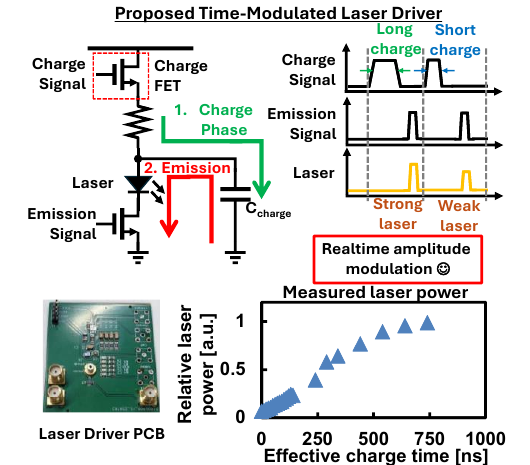}
\caption{Proposed time-modulated laser driver.}
\label{fig:laser}
\end{figure}

To support DDF, microsecond-order laser amplitude modulation is required; however, such modulation cannot be achieved using conventional laser drivers. To address this limitation, this work proposes a time-modulated laser amplitude control scheme that modulates the charge time of a laser capacitor (Fig.~\ref{fig:laser}). In the charge phase, the charge signal turns on the charge FET and charges $C_{\mathrm{charge}}$; in the emission phase, the emission signal discharges $C_{\mathrm{charge}}$ through the laser. A long charge yields a strong laser pulse and a short charge yields a weak laser pulse, enabling real-time amplitude modulation. As shown by the measured laser power in Fig.~\ref{fig:laser}, by controlling the effective charge time from 100 to 1000~ns, the proposed scheme achieves accurate laser amplitude control from 10\% to 100\% with fine resolution.

% ===========================================================================
\section{Implementation}
\label{sec:impl}

\begin{figure}[!t]
\centering
\includegraphics[width=\figw]{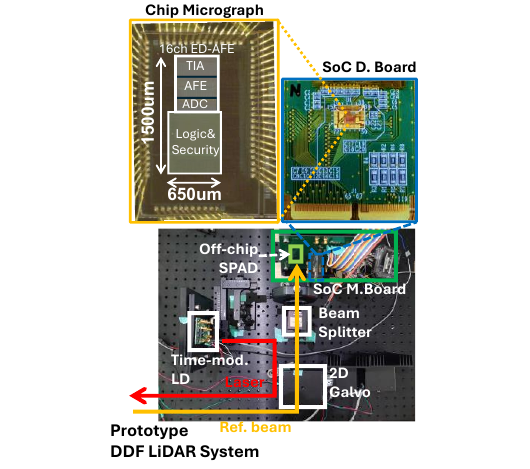}
\caption{Chip photo and prototype DDF LiDAR system.}
\label{fig:chip}
\end{figure}

Fig.~\ref{fig:chip} shows the LiDAR SoC implemented in 65-nm CMOS and the prototype LiDAR system. The 1500~$\mu$m $\times$ 650~$\mu$m chip integrates a 16-channel ED-AFE (TIA, AFE, and ADC) and the logic and security block, and captures off-chip SPAD currents. As shown in Fig.~\ref{fig:arch}, each AFE channel consists of a TIA with DC compensation, event detection and sample-and-hold circuits, and an event ADC; the logic and security block contains the fingerprint authentication and the DDF control that generates the TX modulation for the laser driver. The prototype DDF LiDAR system consists of the SoC on a daughter board and mother board, the time-modulated LD, a beam splitter, a 2-D galvanometer mirror, and the off-chip SPAD.

% ===========================================================================
\section{Measurement Results}
\label{sec:meas}

\subsection{Ranging Performance}

\begin{figure}[!t]
\centering
\includegraphics[width=\figw]{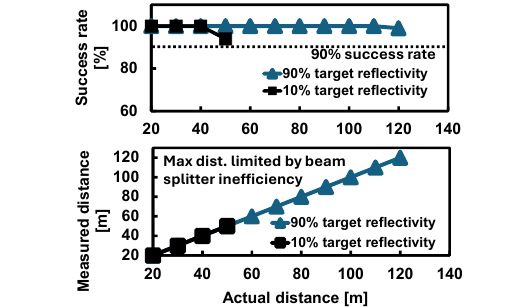}
\caption{Measured ranging performance of the prototype LiDAR: success rate and measured distance versus actual distance for 90\% and 10\% reflectivity targets.}
\label{fig:ranging}
\end{figure}

Fig.~\ref{fig:ranging} shows the LiDAR performance in an indoor long-range setup, achieving a 120-m range with 90\% reflectivity and robust detection for 10\% reflectivity targets up to 50~m. The success rate is evaluated from 1000 trials per target position; a trial is counted as successful when the measured distance lies within 1\% of the actual distance. The maximum distance is limited by the inefficiency of the beam splitter in the optical setup. With its threshold-based filtering, the ED-AFE sustains ranging performance under ambient light up to 30~klux, as confirmed by optical simulations.

\subsection{Resilience Against Spoofing Attacks}

\begin{figure}[!t]
\centering
\includegraphics[width=\figw]{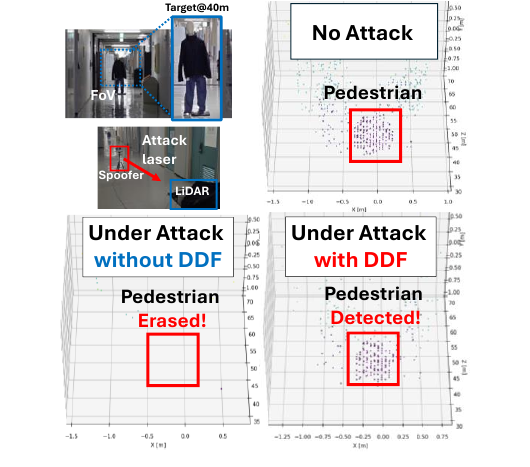}
\caption{Measured LiDAR point clouds under attack (target at 40~m).}
\label{fig:pointcloud}
\end{figure}

The spoofing attack of \cite{Sato_NDSS2024} was applied to the prototype with a spoofer injecting attack laser pulses into the LiDAR, and a pedestrian at 40~m as the target (Fig.~\ref{fig:pointcloud}). In this proof-of-concept experiment, the DDF authentication was performed off-line: the scene was scanned twice under attack with the laser driven at two different drive voltages (80~V and 50~V) to produce two pulse amplitudes, the raw measurement data were captured, and the amplitude fingerprint was applied to the captured data afterwards to emulate the dual-domain authentication.

Table~\ref{tab:attack} summarizes the results. Without an attack, the success rate is 100\% both with and without DDF. Under spoofing attack, all measurements fail when the DDF is disabled. In contrast, enabling the DDF successfully protects 73\% of the point cloud by rejecting spoofed pulses in the 2-D authentication space. Consequently, as shown in Fig.~\ref{fig:pointcloud}, the pedestrian is erased from the point cloud under attack without DDF, but is correctly acquired with DDF.

\begin{table}[!t]
\centering
\caption{Measured Success Rate With and Without DDF}
\label{tab:attack}
\begin{tabular}{lcc}
\toprule
 & w/ DDF & w/o DDF \\
\midrule
Success rate, no attack & 100\% & 100\% \\
Success rate, under attack & \textbf{73\%} & 0\% \\
\bottomrule
\end{tabular}
\end{table}

\subsection{Comparison}

Table~\ref{tab:comparison} compares the proposed LiDAR SoC with prior work using off-chip photodetectors. This work is the first LiDAR SoC in the comparison that targets accurate measurements under active attack, enabled by DDF, and the ED-AFE reduces the AFE power consumption to 3.1~mW/ch, 60\% lower than the 7.8~mW/ch of \cite{Zou_ISSCC2024}.

\begin{table*}[!t]
\centering
\caption{Comparison With State-of-the-Art LiDARs Using Off-Chip Photodetectors}
\label{tab:comparison}
\renewcommand{\arraystretch}{1.1}
\begin{tabular}{lcccccc}
\toprule
 & \textbf{This work} & ISSCC'18 \cite{Yoshioka_ISSCC2018} & ISSCC'20 \cite{Kondo_ISSCC2020} & ISSCC'24 \cite{Zou_ISSCC2024} & ISSCC'20 \cite{Kashmiri_ISSCC2020} & TCAS-I'20 \cite{Liu_TCAS2020} \\
\midrule
Technology & 65~nm & 28~nm & 28~nm & 65~nm & 130~nm & 65~nm \\
ToF method & \textbf{DDF} & Simple dToF & Simple dToF & Simple dToF & Pulse coded & Simple dToF \\
Photodetector & Off-chip & Off-chip & Off-chip & Off-chip & Off-chip & Off-chip \\
Channels & 16 & 20 & 40 & 8 & 1 & 1 \\
Distance range [m] & 50--120$^{\dagger}$ & 200 & 225 & 160--240 & 60 & 30 \\
Laser wavelength [nm] & 905 & 905 & 905 & 905 & 905 & 905 \\
AFE power [mW/ch] & \textbf{3.1} & 39 & 10 & 7.8 & 171 & 7.2 \\
Attack resilient? & \textbf{Yes} & No & No & No & No & No \\
\bottomrule
\multicolumn{7}{l}{\footnotesize $^{\dagger}$Maximum range for 10\% / 90\% reflectivity targets, limited by beam-splitter inefficiency.}
\end{tabular}
\end{table*}

% ===========================================================================
\section{Limitations and Future Work}
\label{sec:limit}

The under-attack evaluation applied the amplitude fingerprint off-line to data captured in two scans at different laser drive voltages; real-time on-chip authentication with pulse pairs generated by the time-modulated laser driver within a single shot remains to be demonstrated. In addition, 27\% of the measurements are still lost under attack, only the high-frequency spoofing attack of \cite{Sato_NDSS2024} was evaluated, and attackers that relay the victim's own pulses would require complementary defenses. The evaluation was carried out indoors with an off-chip photodetector; outdoor evaluation and on-chip SPAD integration are future work.

% ===========================================================================
\section{Conclusion}
\label{sec:conclusion}

This paper presented a secure dToF LiDAR SoC with sensor-level hardware security against spoofing attacks. Dual-Domain Fingerprinting authenticates received laser pulses in both the time and amplitude domains, rejecting attack pulses that merely match the pulse interval. The event-driven AFE activates the ADC only around pulse peaks, reducing ADC power by 99\% while achieving 1-cm distance resolution by parabolic interpolation, and the time-modulated laser driver provides 10\%--100\% laser amplitude modulation. The 16-channel 65-nm CMOS prototype achieves up to 120-m ranging with 3.1~mW/ch AFE power, 60\% lower than prior art, and the proof-of-concept experiment shows 73\% point-cloud protection under attack, compared with 0\% without DDF.

\section*{Acknowledgment}
This research was supported in part by JST CREST JPMJCR23M4, JST PRESTO JPMJPR22PA, JST Next-generation Edge AI Semiconductor JPMJES2515, and JSPS Kakenhi 24K02940.

% ===========================================================================

\end{document}